\documentclass[prd,superscriptaddress,amsfonts,amssymb,amsmath,showpacs,twocolumn]{revtex4-2}
\usepackage{bm}
\usepackage{amsfonts}
\usepackage{latexsym}
\usepackage{graphicx}
\usepackage{amsmath}
\usepackage{palatino}
\usepackage{mathpazo}
\usepackage{textcomp}
\usepackage{float}
\usepackage{booktabs}
\usepackage{dcolumn}
\usepackage{booktabs}
\usepackage{multirow}
\usepackage{hyperref}
\hypersetup{colorlinks,citecolor=blue}
\usepackage{amsmath}
\usepackage{xcolor}
\usepackage{orcidlink}
\usepackage[caption=false]{subfig}
\usepackage{commath}
\def\jnl@style{\it}
\def\aaref@jnl#1{{\jnl@style#1}}

\def\aaref@jnl#1{{\jnl@style#1}}

\def\aj{\aaref@jnl{AJ}}                   
\def\apj{\aaref@jnl{ApJ}}                 
\def\apjl{\aaref@jnl{ApJ}}                
\def\apjs{\aaref@jnl{ApJS}}               
\def\apss{\aaref@jnl{Ap\&SS}}             
\def\aap{\aaref@jnl{A\&A}}                
\def\aapr{\aaref@jnl{A\&A~Rev.}}          
\def\aaps{\aaref@jnl{A\&AS}}              
\def\mnras{\aaref@jnl{Mon.~Not.~Roy.~Astron.~Soc.}}             
\def\prd{\aaref@jnl{Phys.~Rev.~D}}        
\def\prc{\aaref@jnl{Phys.~Rev.~C}}  
\def\prl{\aaref@jnl{Phys.~Rev.~Lett.}}    
\def\qjras{\aaref@jnl{QJRAS}}             
\def\skytel{\aaref@jnl{S\&T}}             
\def\ssr{\aaref@jnl{Space~Sci.~Rev.}}     
\def\zap{\aaref@jnl{ZAp}}                 
\def\nat{\aaref@jnl{Nature}}              
\def\aplett{\aaref@jnl{Astrophys.~Lett.}} 
\def\apspr{\aaref@jnl{Astrophys.~Space~Phys.~Res.}} 
\def\physrep{\aaref@jnl{Phys.~Rep.}}      
\def\physscr{\aaref@jnl{Phys.~Scr}}       
\def\commat{\aaref@jnl{Comm.~Math.~Phys.}}              
\def\science{\aaref@jnl{Science}}               
\def\cqg{\aaref@jnl{Classical Quant.~Grav.}}            
\def\jpcs{\aaref@jnl{JPCS}}                                     
\def\ijmpd{\aaref@jnl{Int.~J.~Mod.~Phys.~D}}                    
\def\grg{\aaref@jnl{Gen.~Relat.~Gravit.}}               
\def\rpp{\aaref@jnl{Rep.~Prog.~Phys.}}          
\def\npa{\aaref@jnl{Nucl.~Phys.~A}}        
\def\lrr{\aaref@jnl{Living Rev.~Rel.}}                   
\def\jcap{\aaref@jnl{J.~Cosmology Astropart.~Phys.}}    
\def\rmp{\aaref@jnl{Rev.~Mod.~Phys.}}   
\def\epjc{\aaref@jnl{Eur.~Phys.~J.~C}}

\allowdisplaybreaks[1]

\begin{document}

\color{black}       

\title{Square-Root parametrization of dark energy in $f\left( Q\right) $ cosmology}

\author{M. Koussour\orcidlink{0000-0002-4188-0572}}
\email[Email: ]{pr.mouhssine@gmail.com}
\affiliation{Quantum Physics and Magnetism Team, LPMC, Faculty of Science Ben
M'sik,\\
Casablanca Hassan II University,
Morocco.} 

\author{N. Myrzakulov\orcidlink{0000-0001-8691-9939}}
\email[Email: ]{nmyrzakulov@gmail.com}
\affiliation{L. N. Gumilyov Eurasian National University, Astana 010008,
Kazakhstan.}
\affiliation{Ratbay Myrzakulov Eurasian International Centre for Theoretical
Physics, Astana 010009, Kazakhstan.}

\author{Alnadhief H. A. Alfedeel\orcidlink{0000-0002-8036-268X}}%
\email[Email:]{aaalnadhief@imamu.edu.sa}
\affiliation{Department of Mathematics and Statistics, Imam Mohammad Ibn Saud Islamic University (IMSIU),\\
Riyadh 13318, Saudi Arabia.}
\affiliation{Department of Physics, Faculty of Science, University of Khartoum, P.O. Box 321, Khartoum 11115, Sudan.}
\affiliation{Centre for Space Research, North-West University, Potchefstroom 2520, South Africa.}

\author{E. I. Hassan\orcidlink{0000-0000-0000-0000}}%
\email[Email:]{eiabdalla@imamu.edu.sa}
\affiliation{Department of Mathematics and Statistics, Imam Mohammad Ibn Saud Islamic University (IMSIU),\\
Riyadh 13318, Saudi Arabia.}

\author{D. Sofuo\u{g}lu\orcidlink{0000-0000-0000-0000}}
\email[Email: ]{degers@istanbul.edu.tr}
\affiliation{Department of Physics, Istanbul University Vezneciler 34134, Fatih, Istanbul, Turkey.}

\author{Safa M.Mirgani\orcidlink{0000-0000-0000-0000}}
\email[Email: ]{smmmohamed@imamu.edu.sa}
\affiliation{Department of Mathematics and Statistics, Imam Mohammad Ibn Saud Islamic University (IMSIU),\\
Riyadh 13318, Saudi Arabia.}
\date{\today}

\begin{abstract}
This paper is a parametrization of the equation of state (EoS) parameter of dark energy (DE), which is parameterized using Square-Root (SR) form i.e. $\omega _{SR}=\text{$\omega _{0}$}+\text{$\omega _{1}$}\frac{z}{\sqrt{z^{2}+1}}$, where $\omega _{0}$ and $\omega _{1}$ are free
constants. This parametrization will be examined in the context of the recently suggested $f(Q)$ gravity theory as an alternative to General Relativity (GR), in which gravitational effects are attributed to the
non-metricity scalar $Q$ with the functional form $f(Q)=Q+\alpha Q^{n}$, where $\alpha$ and $n$ are arbitrary
constants. We derived observational constraints on model parameters using the Hubble dataset with 31 data points and the Supernovae (SNe) dataset from the Pantheon samples compilation dataset with 1048 data points. For the current model, the evolution of the deceleration parameter, density parameter, EoS for DE, and $Om(z)$ diagnostic have all been investigated. It has been shown that the deceleration parameter favors the current accelerated expansion phase. It has also been shown that the EoS parameter for DE has a quintessence nature at this time.
\end{abstract}

\maketitle

\section{Introduction}

\label{sec1}

Observational evidence for high redshift Supernovae (SNe) supports the
growing idea of late-time cosmic acceleration \cite{Riess,Perlmutter}. This
observation is further supported by evidence of Baryon Acoustic Oscillations
(BAO) \cite{D.J.,W.J.}, Cosmic Microwave Background (CMB) \cite{R.R.,Z.Y.},
and Large Scale Structure (LSS) \cite{T.Koivisto,S.F.}. One of the most
delicate difficulties in current cosmology is determining who is accountable for late time cosmic accelerated expansion. Our Cosmos/Universe is governed by an unidentified type of energy known as \textit{Dark Energy}
(DE). Although the inclusion of DE as the cosmological constant, has been
exceedingly efficient, theoretical difficulties of fine-tuning and cosmic
coincidence have hampered its effectiveness \cite{dalal/2001, weinberg/1989}%
. This leads physicists to describe the Cosmos with a phase transition
from deceleration to acceleration. The kinematic method is explored by
Turner and Riess to describe cosmic acceleration without assuming the
validity of general relativity (GR) \cite{Riess1}. This method has no
influence on the physical or geometrical features of DE and is referred to
as \textit{the model-independent way} to investigate DE, i.e. via a
parametrized equation of state (EoS) parameter of DE as a function of scale
factor or redshift, and then comparing such parametrizations to cosmological
data. Ref. \cite{Eric} is a review of the parametrization of the EoS
parameter $\omega \left( z\right) $. Another model-independent way to
investigate the DE is to parametrize the deceleration parameter. For a quick
overview of the deceleration parameter, see here \cite{Yu1}. Nojiri and
Odintsov also investigated numerous parametrizations of the Hubble parameter
to study future cosmological singularities \cite{Nojiri}. The advantage of
the parameterization method is that the result is independent of any specific
gravitational theory. The negative is that it will not provide much direct
knowledge on what is causing the Cosmos to accelerate.

Another approach to solving the late time acceleration problem and
describing the origin of DE is to modify the action of GR, which is known as
Modified theories of gravity (MTG). Many modified theories have been
suggested up to this moment. The $f(R)$ gravity (where $R$ is the scalar
curvature) introduced by Buchdahl is the most important and extensively
utilized modification to GR \cite{Buchdahl}. Numerous researchers have
examined various aspects of $f(R)$ gravity and how it might induce cosmic
inflation and acceleration \cite{fR1, fR2, fR3}. Another extension of the
Einstein-Hilbert action is the presence of non-minimal interaction between
matter and geometry. Thus, the so-called $f(R,T)$ modified theory of gravity
emerges. Harko et al. introduced $f(R,T)$ gravity theory, where the gravitational
Lagrangian is characterized by an arbitrary function of the scalar curvature $R$
and the trace of the energy-momentum tensor $T$ \cite{Harko}. In $f(R,T)$
gravity, several astrophysical and cosmological consequences are studied 
\cite{fRT1, fRT2, fRT3, fRT4, fRT5}. Apart from curvature, the essential
items involved with the manifold's connection defining gravity are also
torsion and non-metricity \cite{Jim1}. The gravity theories may be divided
into three categories based on the connection used. The first employs
curvature, free torsion, and metric-compatible connections, such as GR. The
second class employs a metric-compatible, curvature-free connection with
torsion, such as the teleparallel equivalent of GR \cite{Aldrovandi}. The latter
employs a curvature and torsion-free connection that is not metric
compatible, for example, the Symmetric Teleparallel (ST) equivalent of GR \cite%
{Nester}. The Geometrical Trinity of Gravity refers to these three
equivalent interpretations based on the three separate connections \cite%
{Jim1}. The $f(Q)$ gravity (where $Q$ is the non-metricity scalar) is a
generalization of the ST equivalent of GR with zero torsion and curvature 
\cite{Jim2}. Much research on $f(Q)$ gravity has recently been published.
Refs. \cite{Jim3, Jim4} include the very first cosmological solutions in $f(Q)$
theory, whereas Refs. \cite{Jim5, Jim6} contain $f(Q)$ cosmography and
energy conditions. A power-law model has been examined using quantum
cosmology \cite{Jim7}. Cosmological solutions and matter perturbation growth
index have been examined for a polynomial form of $f(Q)$ theory \cite%
{Jim8}. Harko et al. used a power-law function to analyze the coupling
matter in $f(Q)$ gravity \cite{Jim9}.

In this paper, we consider the Square-Root (SR) parametrization of $\omega
_{DE}\left( z\right) $, and then we use observational data to determine the
behavior of the EoS parameter for DE. After confirming that the Cosmos
underwent acceleration, we presume that the acceleration was caused by DE
and employ a basic DE parametrization to investigate the property of DE
within the framework of $f(Q)$ gravity. The observational constraints on the
model parameters are achieved by utilizing the most recent Hubble dataset
with 31 data points, and SNe dataset from Pantheon samples compilation dataset with 1048 data points. Using the estimated values of model
parameters, we examined the evolution of the deceleration parameter and the
EoS parameter for DE at the $1-\sigma $ and $2-\sigma $ confidence levels.
This study is structured as follows: in the next section, we describe the
fundamental cosmological scenario of $f(Q)$ gravity and derive the field
equations in for flat FRW space. Sec. \ref{sec3} presents the
parametrization model of the EoS parameter and the background for discussing
the cosmic evolution of the Cosmos. In Sec. \ref{sec4}, we have detailed
the most recent observational data-sets used in our research. Sec. \ref{sec5}
goes with cosmological parameters such as deceleration parameter, energy
density, EoS parameter, and $Om(z)$ diagnostic. The last section
contains the outcomes.

\section{$f(Q)$ cosmology}

\label{sec2}

In $f(Q)$ theory, the covariant derivative of the metric tensor is non-zero,
and this fact can be represented mathematically in terms of a new geometrical variable known as non-metricity i.e. $Q_{\sigma \mu \nu }=\nabla _{\sigma
}g_{\mu \nu }\,$. So, $f(Q)$ gravity is a modified theory that extends
Einstein's theory of GR by adding a function of the non-metricity tensor $Q$
into the action, which has no curvature or torsion. The non-metricity tensor
measures the variation of a vector's length in parallel transport and is the
critical geometric variable that explains the characteristics of
gravitational interaction. The action of the $f(Q)$ gravity is \cite{Jim2} 
\begin{equation}
S=\int \left[ -\frac{1}{2}f(Q)+\mathcal{L}_{m}\right] \sqrt{-g}~d^{4}x,
\label{qqm}
\end{equation}%
where $Q$ is replaced by $f(Q)$ in the symmetric teleparallel action, $f(Q)$
being an arbitrary function of $Q$. Here, $g$ is the determinant of the
metric tensor $g_{\mu \nu }$ and $\mathcal{L}_{m}$ is the usual matter
Lagrangian density. The two independent traces of $Q_{\alpha \mu \nu }$ are 
\begin{equation}
Q_{\sigma }=Q_{\sigma }{}^{\mu }{}_{\mu }\,,\quad \tilde{Q}_{\sigma }=Q^{\mu
}{}_{\sigma \mu }\,.
\end{equation}

Moreover, the non-metricity scalar is defined as a contraction of $Q_{\alpha
\mu \nu }$ given by 
\begin{equation}
Q=-Q_{\sigma \mu \nu }P^{\sigma \mu \nu }\,,
\end{equation}%
where $P^{\sigma \mu \nu }$ is the superpotential tensor (also known as the
non-metricity conjugate) and 
\begin{equation}
4P^{\sigma }{}_{\mu \nu }=-Q^{\sigma }{}_{\mu \nu }+2Q_{(\mu \phantom{\sigma}%
\nu )}^{\phantom{\mu}\sigma }-Q^{\sigma }g_{\mu \nu }-\tilde{Q}^{\sigma
}g_{\mu \nu }-\delta _{(\mu }^{\sigma }Q_{\nu )}\,.  \label{super}
\end{equation}

A variation of action (\ref{qqm}) with regard to the metric gives the field
equations as 
\begin{widetext}
\begin{equation}
\frac{2}{\sqrt{-g}}\nabla _{\sigma }\left( \sqrt{-g}f_{Q}P^{\sigma }{}_{\mu
\nu }\right) +\frac{1}{2}g_{\mu \nu }f+f_{Q}\left( P_{\mu \sigma \beta
}Q_{\nu }{}^{\sigma \beta }-2Q_{\sigma \beta \mu }P^{\sigma \beta }{}_{\nu
}\right) =T_{\mu \nu }\,,  \label{EFE}
\end{equation}%
\end{widetext}where $f_{Q}=f_{Q}\left( Q\right) =\frac{df\left( Q\right) }{dQ%
}$ and $T_{\mu \nu }=-\frac{2}{\sqrt{-g}}\frac{\delta \left( \sqrt{-g}%
\mathcal{L}_{m}\right) }{\delta g^{\mu \nu }}$ with a choice of unit as $%
8\pi G=c=1$.

We consider that the matter of the Cosmos is a perfect fluid with no
viscosity. The energy-momentum tensor $T_{\mu \nu }$ is given by 
\begin{equation}
\label{CE}
T_{\mu \nu }=(\rho +p)u_{\mu }u_{\nu }+pg_{\mu \nu }\,,
\end{equation}%
where $u_{\mu }$ is the 4-velocity satisfying the normalization condition $%
u_{\mu }u^{\mu }=-1$. Also, $\rho $ and $p$ are the energy density and
isotropic pressure of a perfect fluid, respectively. The current objective
is to investigate the dynamics of the Cosmos against the background of the
spatially flat Friedmann-Lemaitre-Robertson-Walker (FLRW) metric, which is
expressed as 
\begin{equation}
ds^{2}=-dt^{2}+a^{2}(t)\left[ dr^{2}+r^{2}\left( d\theta ^{2}+\sin
^{2}\theta d\phi ^{2}\right) \right] .
\end{equation}

In this case, the non-metricity scalar is given by $Q=6H^{2}$, where $H=%
\frac{\dot{a}}{a}$ is the Hubble parameter with $a(t)$ as the scale factor
and the dot is the derivative with respect to cosmic time $t$. The modified
Friedmann equations govern the Cosmos's dynamics take the form \cite{Jim3}%
\begin{equation}
3H^{2}=\frac{1}{2f_{Q}}\left( \rho+\frac{f}{2}\right) ,  \label{F1}
\end{equation}%
\begin{equation}
\dot{H}+\left( 3H+\frac{\dot{f_{Q}}}{f_{Q}}\right) H=\frac{1}{2f_{Q}}\left(
-p+\frac{f}{2}\right) .  \label{F2}
\end{equation}

For $f(Q)=Q$, we get the standard Friedmann equations \cite{Jim3}, as
predicted, because, as previously stated, this specific option for the form of the function $f(Q)$ represents the theory's ST equivalent
of GR limit. By using $f(Q)=Q+F(Q)$, the field equations (\ref{F1}) and (\ref%
{F2}) can be written as%
\begin{equation}
3H^{2}=\rho+\frac{F}{2}-QF_{Q}\,,  \label{F11}
\end{equation}%
\begin{equation}
\left( 2QF_{QQ}+F_{Q}+1\right) \dot{H}+\frac{1}{4}\left( Q+2QF_{Q}-F\right)
=-2p\,.  \label{F22}
\end{equation}%

Eqs. \eqref{F11} and \eqref{F22} can be interpreted as the ST counterparts of GR cosmology, but with the inclusion of an additional component arising from the non-metricity $Q$ of space-time that exhibits properties similar to those of a fluid component attributed to DE i.e. $\rho_{Q}=\rho_{DE}$ and $p_{Q}=p_{DE}$.

Therefore, using Eqs. \eqref{F11} and \eqref{F22}, we obtain%
\begin{equation}
H^{2}=\frac{1}{3}\left( \rho+\rho _{DE}\right) \,,
\end{equation}%
\begin{equation}
2\dot{H}+3H^{2}=-(p+p_{DE})\,,
\end{equation}%
where $\rho _{DE}$ and $p_{DE}$ are the density and pressure contributions
of the DE due to the non-metricity of space-time defined by%
\begin{equation}
\rho _{DE}=\frac{F}{2}-QF_{Q}\,,  \label{F111}
\end{equation}%
\begin{equation}
p_{DE}=2\dot{H}(2QF_{QQ}+F_{Q})-\rho _{DE}\,.  \label{F222}
\end{equation}

Furthermore, the equation of state (EoS) parameter due to the DE component is 
\begin{equation}
\omega _{DE}=\frac{p_{DE}}{\rho _{DE}\,}=-1+\frac{4\dot{H}(2QF_{QQ}+F_{Q})}{%
F-2QF_{Q}}\,.
\end{equation}

By applying the covariant divergence to the field equations (\ref{EFE}), we derive $\nabla^{\mu} T_{\mu  \nu}$=0  \cite{Avik}, and incorporating Eq. (\ref{CE}), the conservation equation for the energy-momentum tensor can be derived as
\begin{equation}
\dot{\rho}+3H(1+\omega)\rho=0
\end{equation}

\section{Square-Root (SR) parametrization for EoS parameter}

\label{sec3}

For our investigation of SR parametrization, we pressume the functional form 
$F\left( Q\right) =\alpha Q^{n}$, where $\alpha $ and $n$ are model
parameters. This specific functional form of $f(Q)$ is motivated in Ref. 
\cite{Jim4}. Also, we observe that if $n=0$, the model reduces to the
standard $\Lambda $CDM model, with $\frac{\alpha }{2}$ behaving as the
cosmological constant \cite{Jim10, Jim11}. The scenario $n=1$ corresponds to
the Symmetric Teleparallel Equivalent of GR, due to a factor of $\alpha +1$
rescaling of Newton's gravitational constant \cite{Jim3}. Nevertheless,
modification from the GR evolution occurs in the small curvature phase for $%
n<1$ and modification at the large curvature phase for $n>1$. Thus, whereas
models with $n>1$ will apply to the early Cosmos, models with $n<1$ will
apply to the late time DE-dominated Cosmos \cite{Jim4}.

Using this form, we derived the energy density $\rho _{DE}$, and pressure $%
p_{DE}$ for DE in terms of Hubble parameter as, 
\begin{equation}
\rho_{DE} =\alpha6^n(\frac{1}{2}- n) H^{2n},  \label{DE1}
\end{equation}
and
\begin{equation}
p_{DE}=-\alpha  6^{n-1} (\frac{1}{2}- n) H^{2(n-1)} \left(3 H^2+2n \dot{H}\right).  \label{DE2}
\end{equation}

The EoS parameter $\omega _{DE}$ for DE becomes%
\begin{equation}
\omega_{DE} =-1-\frac{2n}{3}\left( \frac{\overset{.}{H}}{H^{2}}\right).  \label{DE3}
\end{equation}

To get cosmological findings that enable for direct comparison of predicted results with observational data, we include the redshift parameter $z$,
expressed by $1+z=\frac{1}{a}$, as an independent variable instead of the
cosmic time variable $t$. Here, we have normalized the scale factor so that
its current day value is one i.e. $a\left( 0\right) =a_{0}=1$. Therefore,
the time-dependent derivative of the Hubble parameter is represented as%
\begin{equation}
\overset{.}{H}=\frac{dH}{dt}=-\left( 1+z\right) H\left( z\right) \frac{%
dH\left( z\right) }{dz}.  \label{Hd}
\end{equation}

Now, we have an additional choice to pick one parameter because we have more
unknown parameters with fewer equations to solve i.e. Eqs. (\ref{DE1})-(\ref%
{DE3}). The DE is typically described by an EoS parameter that is a ratio of
spatially homogenous pressure to DE density. Recent cosmological
investigations show that the ambiguities are too great to distinguish
between the cases: $\omega <-1$, $\omega =-1$, and $\omega >-1$ \cite{WMAP9,
Planck2013, Planck2020}. In the decelerating phase, which contains two
epochs of matter and radiation, the EoS parameter takes the values $\omega =%
\frac{1}{3}$ and $\omega =0$ respectively. The accelerating phase of the
Cosmos, which has recently been explored, is represented with $\omega <-%
\frac{1}{3}$ which contains the quintessence $-1<\omega <-\frac{1}{3}$, $%
\Lambda $CDM $\omega =-1$, and phantom regime $\omega <-1$. The value of the
EoS parameter for DE estimated by Nine-Year Wilkinson Microwave Anisotropy
Probe (WMAP9), which integrated data from $H_{0}$ measurements, SNe, CMB,
and BAO, is $\omega =-1.084\pm 0.063$ \cite{WMAP9}, whereas the Planck
collaboration shows $\omega =-1.006\pm 0.0451$ \cite{Planck2013}, and later
in 2018, it published $\omega =-1.028\pm 0.032$ \cite{Planck2020}. So, the
value of the EoS parameter flips from positive in the past to negative in
the present. The parameterization of the EoS is a valuable way for
reconstructing cosmological parameters and constraining the dynamical
history of the Cosmos in a general scheme. There are numerous
parameterizations for $\omega _{DE}$ described in the literature, see Refs. 
\cite{BA, Nunes, Wei, Jaime}. In this work, we consider the SR
parametrization form of EoS parameter for DE in terms of redshift $z$ ($%
\omega _{Q}=\omega _{DE}=\omega _{SR}$) \cite{RS},%
\begin{equation}
\omega _{SR}=\text{$\omega _{0}$}+\text{$\omega _{1}$}\frac{z}{\sqrt{z^{2}+1}%
}.  \label{DE4}
\end{equation}%

The parametrization employed in this study offers several key advantages in the context of modeling the EoS of DE. Firstly, it provides a physically interpretable framework, with $\omega_0$ representing the DE EoS at the present epoch ($z=0$) and $\omega_1$ characterizing the temporal evolution of DE. This transparency enhances our understanding of the underlying physical processes. Moreover, the parametrization explicitly incorporates the redshift dependence, allowing us to capture the evolving nature of DE over cosmic time.  Its flexibility enables us to encompass a wide range of DE behaviors, from quintessence to phantom DE. This adaptability ensures that the parametrization can be fine-tuned to match a variety of observational data, making it a valuable tool for cosmological research. In addition, it clearly
observes that the EoS parameter at high redshift (i.e. $z\rightarrow \infty $%
) becomes $\omega _{SR}=\omega _{0}$$+\omega _{1}$ and thus depends on the
values $\omega _{0}$ and $\omega _{1}$. For $z\rightarrow -1$,
becomes as $\omega _{SR}=\omega _{0}-\frac{\omega _{1}}{\sqrt{2}}$, it
behaves like DE throughout cosmic evolution and this is appropriate for $%
\omega _{DE}$ in Eq. (\ref{DE3}).

Using Eqs. (\ref{DE3}), (\ref{Hd}), and (\ref{DE4}), we obtain the following
differential equation%
\begin{equation}
\frac{dH\left( z\right) }{dz}=\frac{3\left( \text{$\omega _{0}$}+\frac{\text{%
$\omega _{1}$}z}{\sqrt{z^{2}+1}}+1\right) }{2n(z+1)}H\left( z\right)
\label{DE5}
\end{equation}

Solving Eq. (\ref{DE5}), we obtain the expression for the Hubble parameter in
terms of $z$%
\begin{widetext}
\begin{equation}
H\left( z\right) =c_{0}\exp \left( \frac{3\left( \frac{\text{$\omega _{1}$}%
\tanh ^{-1}\left( \frac{1-z}{\sqrt{2}\sqrt{z^{2}+1}}\right) }{\sqrt{2}}+(%
\text{$\omega _{0}$}+1)\log (z+1)+\text{$\omega _{1}$}\sinh ^{-1}(z)\right) 
}{2n}\right) ,  \label{DE6}
\end{equation}%
\end{widetext}
where $c_{0}$ is a constant of integration. The current value of the Hubble
parameter can be obtained as%
\begin{equation}
H_{0}=H\left( z=0\right) =c_{0}e^{\frac{3\text{$\omega _{1}$}\coth
^{-1}\left( \sqrt{2}\right) }{2\sqrt{2}n}}.  \label{DE7}
\end{equation}

Using Eqs. (\ref{DE6}) and (\ref{DE7}), the Hubble parameter can be
rewritten in terms of $H_{0}$ in the form%
\begin{widetext}
\begin{equation}
H\left( z\right) =H_{0}(z+1)^{\frac{3(\text{$\omega _{0}$}+1)}{2n}}\exp
\left( \frac{6\text{$\omega _{1}$}\sinh ^{-1}(z)-3\sqrt{2}\text{$\omega _{1}$%
}\left( \tanh ^{-1}\left( \frac{z-1}{\sqrt{2}\sqrt{z^{2}+1}}\right) +\coth
^{-1}\left( \sqrt{2}\right) \right) }{4n}\right) .
\label{Hz}
\end{equation}
\end{widetext}

The deceleration parameter $q$ is written as%
\begin{equation}
q=-\frac{\overset{..}{a}}{aH^{2}}=-1+\frac{\left( 1+z\right) }{H\left(
z\right) }\frac{dH\left( z\right) }{dz}.
\end{equation}

In the current model, $q$ evolves as a function of $z$ as%
\begin{equation}
q\left( z\right) =-1+\frac{1}{2n}\left( 3\text{$\omega _{0}$}+\frac{3\text{$%
\omega _{1}$}z}{\sqrt{z^{2}+1}}+3\right) .
\end{equation}

The $Om$ diagnostic is an important method for classifying the various DE
cosmological scenarios \cite{Omz}. It is the simplest diagnosis since it
just takes the first order derivative of the cosmic scale factor. It is
expressed for a spatially flat Cosmos as%
\begin{equation}
Om\left( z\right) =\frac{\left( \frac{H\left( z\right) }{H_{0}}\right) ^{2}-1%
}{\left( 1+z\right) ^{3}-1}.
\end{equation}

The negative slope of $Om(z)$ leads to quintessence type behavior $%
(\omega>-1)$, whereas the positive slope, refers to phantom-type behavior $%
(\omega<-1)$. The constant nature of $Om(z)$ represents the $\Lambda $CDM
model $(\omega=-1)$. In the next section, we will attempt to estimate the $%
\omega _{0}$, $\omega _{1}$ and $n$ values using the Hubble and Pantheon
data-sets. With the help of the $\omega _{0}$, $\omega _{1}$ and $n$ values,
we discuss the behavior of the aforementioned cosmological parameters and
verify the validity of the cosmological model.

\section{Observational data}

\label{sec4}

The several observational datasets may then be utilized to restrict the
parameters $H_{0}$, $\omega _{0}$, $\omega _{1}$ and $n$. It is important to note
that the EoS parameter $\omega _{DE}$ for DE (see Eq. (\Ref{DE3})) is not
dependent on the parameter $\alpha $ and hence is not explicitly contained
in the Hubble parameter expression, i.e. Eq. (\ref{Hz}), we attempt to fix
it by setting $\alpha =1$. We study the observational data using the
standard Bayesian approach, and we get the posterior distributions of the
parameters using the MCMC (Markov Chain Monte Carlo) method. In addition,
for MCMC analysis, we use the \textit{emcee} Python package \cite%
{Mackey/2013}. In this work, we use two data: Hubble dataset with 31 data
points, and SNe dataset from Pantheon samples compilation dataset with
1048 data points.

\subsection{Hubble dataset}

We utilize a dataset consisting of 31 data points obtained through the Cosmic Chronometers (CC) technique. This approach allows us to directly extract information about the Hubble function at various redshifts, spanning up to $z \leq 2$. The choice to incorporate CC data is primarily motivated by its reliance on measurements of age differences between two galaxies that evolved passively, originating simultaneously but separated by a small redshift interval. This method facilitates the calculation of $\Delta z/\Delta t$. Notably, CC data has demonstrated higher reliability in comparison to other methods that depend on absolute age determinations for galaxies, as previously discussed in \cite{Jimenez11}. The CC data points utilized in our analysis were collected from references \cite{Zhang1,Jimenez22,Moresco11,Simon1,Moresco22,Stern1,Moresco33}, all of which are independent of the Cepheid distance scale and specific cosmological models. It is important to acknowledge that these data points rely on the modeling of stellar ages, employing well-established techniques of stellar population synthesis (for further details, refer to Refs. \cite{Moresco11,Moresco22,Valent1,Corredoira11,Corredoira22,Verde1} for analyses related to CC systematics). To evaluate the goodness of fit between our model and the dataset, we employ the $\chi^2$ function, defined as
\begin{equation}
\chi _{Hubble}^{2}(H_{0},\omega _{0},\omega _{1},n)=\sum_{i=1}^{57}\frac{%
[H_{i}^{th}(H_{0},\omega _{0},\omega _{1},n,z_{i})-H_{i}^{obs}(z_{i})]^{2}}{\sigma
_{Hubble}^{2}(z_{i})}.  \label{4a}
\end{equation}

Here, $H_{i}^{obs}$ denotes the observed value, $H_{i}^{th}$ is the theoretical value of the Hubble parameter obtained by our model, and $\sigma
_{z_{i}}$ is the standard error in the observed value.

\subsection{Pantheon dataset}

The SNe plays an important role in characterizing the expanding Cosmos.
Moreover, spectroscopically acquired SNe data from studies that include the
SuperNova Legacy Survey (SNLS), the Sloan Digital Sky Survey (SDSS), the
Hubble Space Telescope (HST) survey, and the Panoramic Survey Telescope and
Rapid Response System (Pan-STARRS1) provide strong evidence in this path.
The Pantheon dataset, the most recent SNe data sample, comprise 1048 magnitudes for the distance modulus evaluated over a redshift range of $%
0.01<z<2.3$ \cite{Scolnic/2018, Chang}. For the Pantheon dataset, the $\chi
^{2}$ function is given by, 
\begin{equation}
\chi _{Pan}^{2}(H_{0},\omega _{0},\omega _{1},n )=\sum_{i,j=1}^{1048}\bigtriangledown
\mu _{i}\left( C_{Pan}^{-1}\right) _{ij}\bigtriangledown \mu _{j}.
\label{4b}
\end{equation}

Here, $C_{Pan}$ represents the covariance matrix \cite{Scolnic/2018}, and $%
\bigtriangledown \mu _{i}=\mu ^{th}(z_{i},\theta )-\mu _{i}^{obs}(z_{i})$
represents the difference between the observed distance modulus value
acquired from cosmic measurements and its theoretical values obtained by our
model. The theoretical and observed distance modulus are denoted by $\mu
_{i}^{th}$ and $\mu _{i}^{obs}$, respectively. The theoretical distance
modulus is $\mu _{i}^{th}(z)=m-M=5LogD_{l}(z)$, in which $m$ and $M$ are the
apparent and absolute magnitudes of a standard candle, respectively. In
addition, the luminosity distance $D_{l}(z)$ is \cite{Planck2020},%
\begin{equation}
D_{l}(z)=c(1+z)\int_{0}^{z}\frac{dy}{H(y)}.
\end{equation}

\subsection{Hubble+Pantheon dataset}

In this subsection, we presented the outcomes of the statistical MCMC
method combined with the Bayesian analysis. We apply the joint analysis
for the Hubble dataset with 31 data points, and the Pantheon dataset with
1048 sample points to constrain the parameters of the model $H_0$, $\omega _{0}$, $%
\omega _{1}$ and $n$ from Eq. (\ref{Hz}). We utilized 100 walkers and 1000 MCMC steps to get conclusions for the dataset. For the joint analysis, we consider the following priors: $H _{0}\in \left[60,80\right] $, $\omega _{0}\in \left[ -2,2\right] $, $%
\omega _{1}\in \left[ -2,2\right] $, and $n\in \left[ -2,2\right] $.
When only a dataset is used to estimate parameters, we presume a Gaussian
likelihood. The quality of fit for the joint analysis is measured by a total chi-squared function $\chi _{T}^{2}$ that is defined as:%
\begin{equation}
\chi _{T}^{2}=\chi _{Hubble}^{2}+\chi _{Pan}^{2},
\end{equation}%
where $\chi _{Hubble}^{2}$ is calculated using Eq. (\ref{4a}) and $\chi
_{Pan}^{2}$ is calculated using Eq. (\ref{4b}). Moreover, a joint Gaussian
likelihood can be written as%
\begin{equation}
{\mathcal{L}}_{T}\propto e^{-\frac{\chi _{T}^{2}}{2}},
\end{equation}%
where ${\mathcal{L}}_{T}$ represents the product of the likelihood functions
of each dataset.

\begin{widetext}	

\begin{figure}[H]
\centering
\includegraphics[scale=0.80]{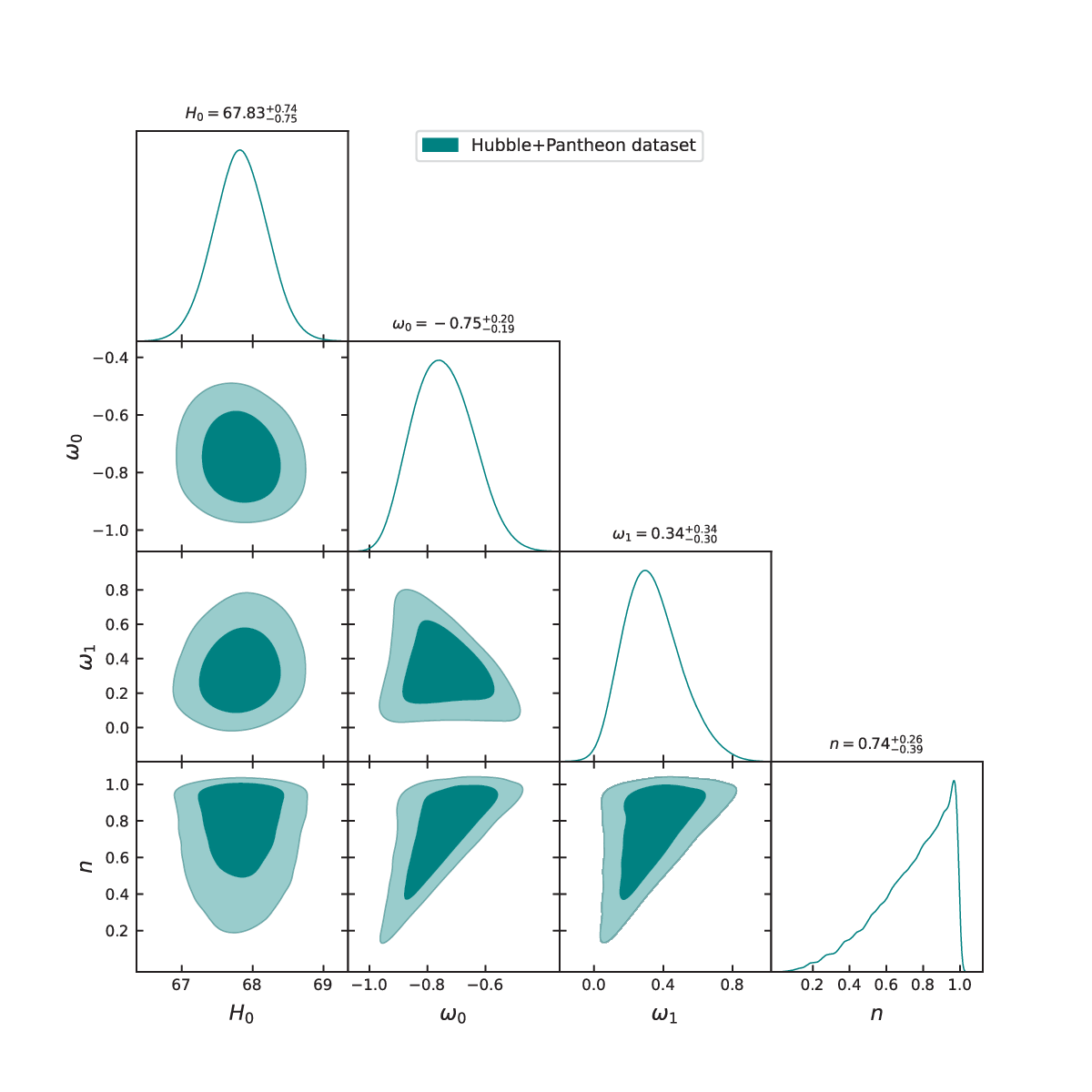}
\caption{The confidence curves for the model parameters at $1-\sigma$ and $2-\sigma$ using the Hubble+Pantheon dataset. Dark green shaded zones indicate the $1-\sigma$ confidence level (CL), whereas light green shaded regions reflect the $2-\sigma$ CL. The parameter constraint values are also displayed at the $1-\sigma$ CL.}
\label{Contour}
\end{figure}	

\begin{figure}[H]
\centering
\includegraphics[scale=0.55]{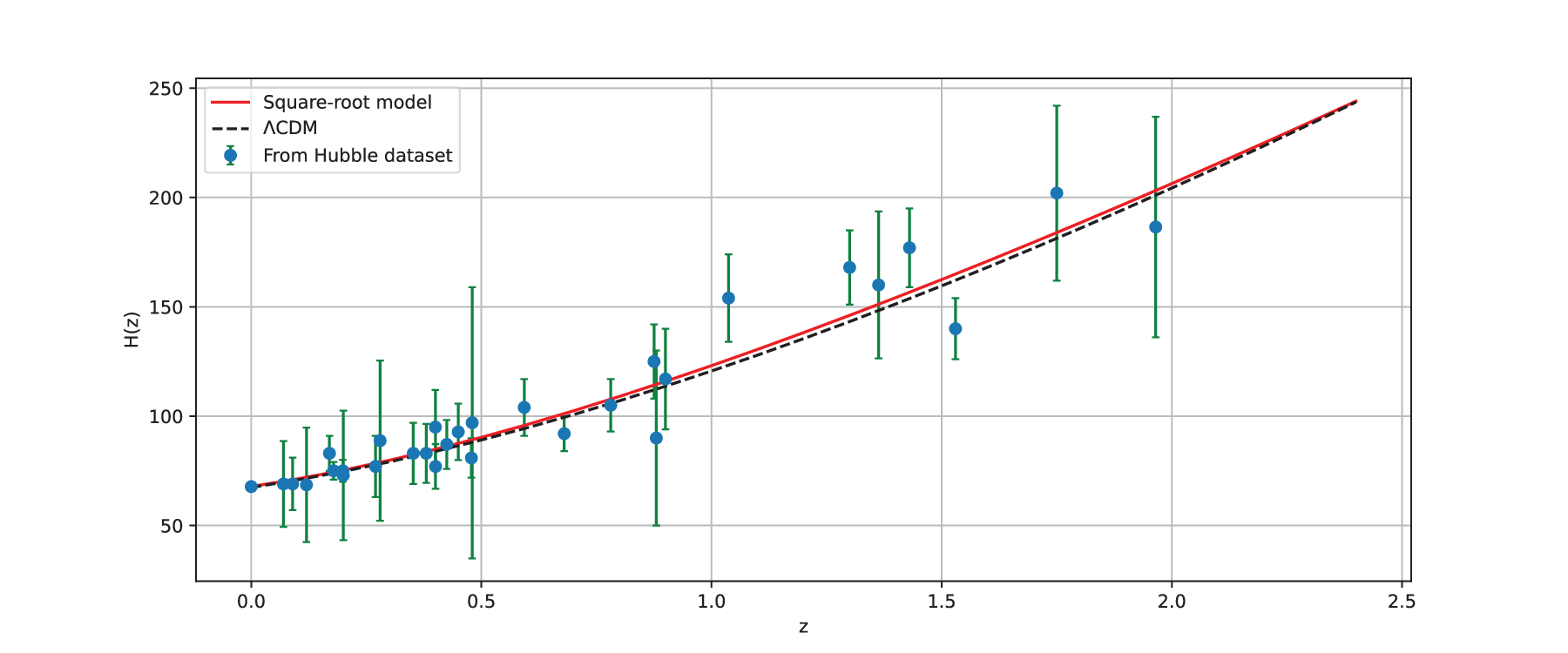}
\caption{The variation of $H(z)$ vs. $z$. The blue dots represent error bars, the red line represents our model's curve, and the black dashed line represents the $\Lambda$CDM model.}
\label{H}
\end{figure}	

\begin{figure}[H]
\centering
\includegraphics[scale=0.55]{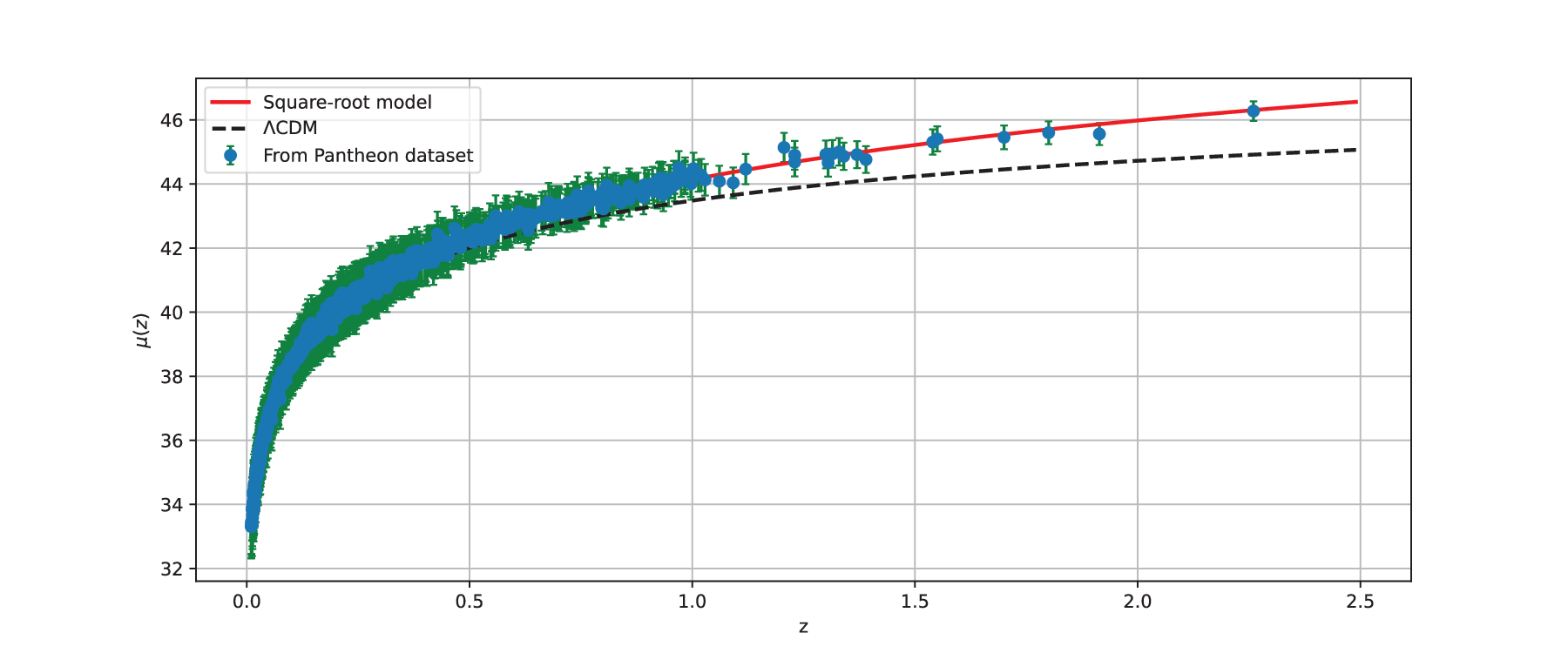}
\caption{The variation of $\mu(z)$ vs. $z$. The blue dots represent error bars, the red line represents our model's curve, and the black dashed line represents the $\Lambda$CDM model.}
\label{Mu}
\end{figure}

\end{widetext}

Recent studies have highlighted the significance of Hubble parameter measurements and Pantheon data in constraining cosmological parameters. In our model, the parameters of interest are $H_{0}$, $\omega _{0}$, $\omega _{1}$, and $n$. However, our primary aim is to assess the influence of parameters within the chosen parameter space and gauge their compatibility with observational data. Fig. \ref{Contour} displays the $1-\sigma$ and $2-\sigma$ likelihood contours derived from the joint analysis of the Hubble and Pantheon datasets. The best-fit values of the estimated model parameters are $H _{0}=67.83_{-0.75}^{+0.74}$, $\omega _{0}=-0.75_{-0.19}^{+0.20}$, $\omega
_{1}=0.34_{-0.30}^{+0.34}$, and $n=0.74_{-0.39}^{+0.26}$. The likelihood functions for the Hubble+Pantheon datasets are also extremely well matched to a Gaussian distribution function, as seen in Fig. %
\ref{Contour}. Figs. \ref{H} and \ref{Mu} compare our RS parametrization model to the widely accepted $\Lambda $CDM model in cosmology, i.e. $H\left(
z\right) =H_0\sqrt{\Omega _{0}^{m}\left( 1+z\right) ^{3}+\Omega _{0}^{\Lambda }}
$. For the figure, we choose $\Omega _{0}^{m}=0.314$ and $H_{0}=67.4$ $km.s^{-1}.Mpc^{-1}$ 
\cite{Planck2020}. The figures also include Hubble and Pantheon experimental
findings, with 31 and 1048 data points with errors, respectively, allowing
for a direct comparison between different models.

\section{Cosmological parameters}

\label{sec5}

\begin{figure}[]
\centering
\includegraphics[scale=0.7]{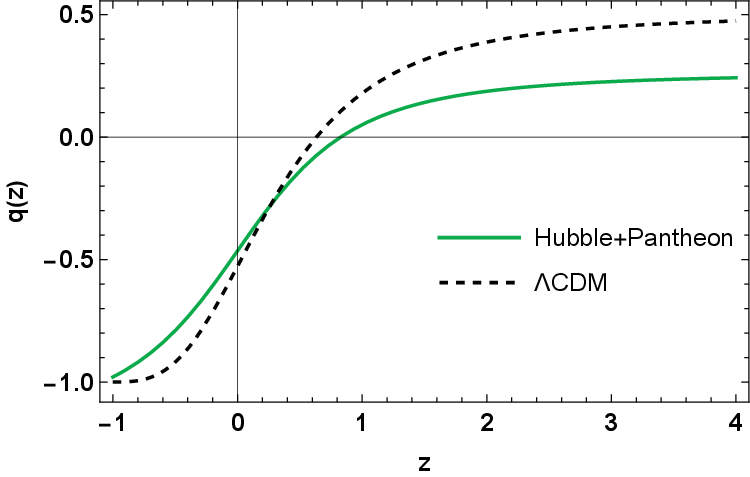}
\caption{The behavior of the deceleration parameter $q$ vs. redshift $z$.}
\label{q}
\end{figure}

\begin{figure}[]
\centering
\includegraphics[scale=0.68]{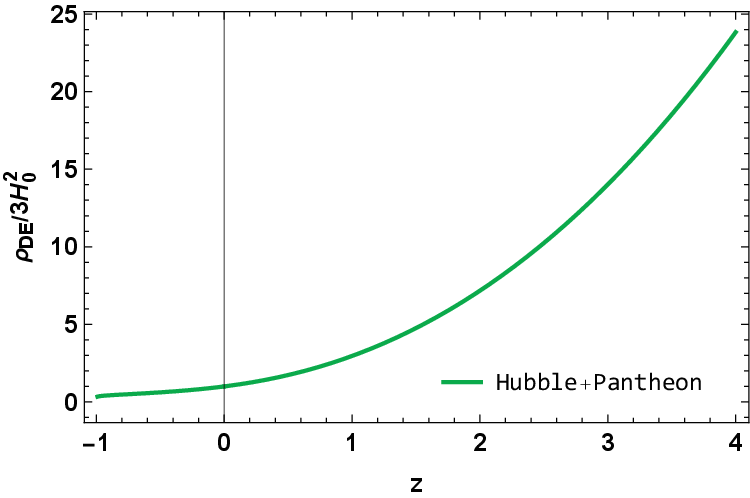}
\caption{The behavior of the density parameter for DE $\rho_{DE}$ vs.
redshift $z$.}
\label{rhoDE}
\end{figure}

Cosmological parameters play a vital role in the construction of
cosmological models. For the model to be realistic, the current value of the
deceleration parameter $q$ must be matched with the cosmic measurements.
Furthermore, the model's decelerating or accelerating behavior is determined
by the positive or negative value of $q$. The behavior of the deceleration
parameter, energy density, and EoS parameter in terms of redshift is
illustrated below. For the model parameter values, we use the
Hubble+Pantheon data-sets. According to Fig. \ref{q}, the deceleration
parameter clearly indicates the transition from a decelerated (i.e., $q>0$)
to an accelerated (i.e., $q<0$) stage of the Cosmos expansion for the
constrained values of the model parameters. The present value of the
deceleration parameter (i.e., $z=0$) corresponding to the model parameter
values constrained by the Hubble+Pantheon dataset is $%
q_{0}=-0.46_{-0.07}^{+0.16}$. Moreover, the transition redshift
(i.e., $q=0$) is $z_{tr}=0.83_{-0.33}^{+0.54}$ \cite{Mamon/2018,Farooq/2017} for the
Hubble+Pantheon dataset. In addition, it is essential to point out that the $q_{0}$ and $z_{tr}$ values constrained in this paper are consistent with the value reported in Refs. \cite{Capozziello, Mamon, Basilakos}. The behavior of the energy density of DE is shown in Fig. \ref{rhoDE}. As the Universe expands, the energy density of DE increases with redshift $z$, but decreases with time. At late times, the DE density reaches a minimum value. Moreover, the small value of DE density implies that the Universe will continue to accelerate in its expansion in the future, leading to the big rip scenario.

\begin{figure}[]
\centering
\includegraphics[scale=0.7]{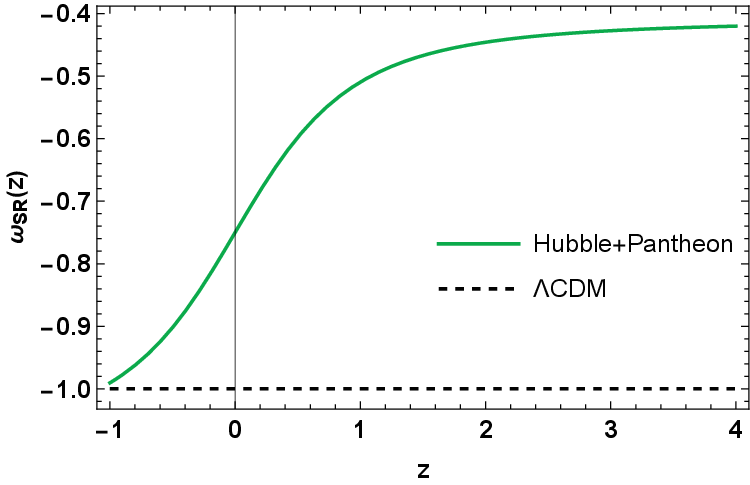}
\caption{The behavior of the EoS parameter for DE $\protect\omega$ vs.
redshift $z$.}
\label{EoS}
\end{figure}

The EoS parameter is one of the cosmological parameters that are important in describing the status of the expansion of our Cosmos. So when the value of the
EoS parameter is strictly less than $\omega <-\frac{1}{3}$, the Cosmos
accelerates. The behavior of the EoS parameter for DE is depicted in Fig. %
\ref{EoS} based on constrained values of model parameters $\omega _{0}$, $%
\omega _{1}$ and $n$ from Hubble+Pantheon data-sets. It is seen that the EoS
parameter for DE continues in the quintessence era, maintaining the
Cosmos's acceleration. The current value of EoS is calculated as $\omega _{0}=-0.75_{-0.19}^{+0.20}$ \cite{Hernandez, Zhang}.

\begin{figure}[]
\centering
\includegraphics[scale=0.7]{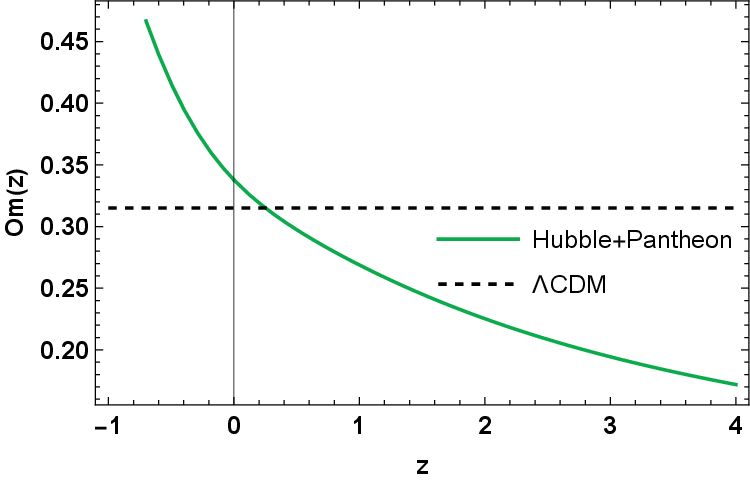}
\caption{The behavior of the $Om(z)$ vs. redshift $z$.}
\label{Om}
\end{figure}

In addition, the slope of the diagnostic parameter $Om(z)$ can distinguish
between two types of DE scenarios (quintessence and phantom). According to
Fig. \ref{Om}, the $Om\left( z\right) $ for the constrained values of the
model parameters has a negative slope over the whole domain. We may deduce
from the $Om\left( z\right) $ diagnostic test that our RS model depicts quintessence-type behavior. This indicates that our model exhibits unique characteristics when contrasted with the standard $\Lambda$CDM model.

\section{Conclusion}

\label{sec6}

DE is one of the most enticing and intriguing topics in current cosmology,
as it contributes to the Cosmos's accelerating expansion. Many attempts
have been taken to explain this cosmic acceleration, including several
parametrizations of DE models and modified theories of gravity. The goal of
this research is to examine the SR parametrization model in the $f(Q)$
theory of gravity as an alternative theory to GR, in which gravitational
effects are attributed to the non-metricity scalar $Q$. We derived the exact
solution of the field equations for the functional form of $f(Q)$ as $%
f(Q)=Q+\alpha Q^{n}$, where $\alpha$ and $n$ are arbitrary constants, by
using the SR parametrization form of the EoS parameter for DE as $\omega
_{SR}=\text{$\omega _{0}$}+\text{$\omega _{1}$}\frac{z}{\sqrt{z^{2}+1}}$,
which leads to the variable deceleration parameter. To obtain the best-fit values for the model parameters $\omega _{0}$, $\omega _{1}$ and $n$, we
used Hubble dataset with 31 data points and SNe dataset from Pantheon
samples compilation dataset with 1048 data points. The best-fit values of
the estimated model parameters are $H _{0}=67.83_{-0.75}^{+0.74}$, $\omega _{0}=-0.75_{-0.19}^{+0.20}$, $\omega
_{1}=0.34_{-0.30}^{+0.34}$, and $n=0.74_{-0.39}^{+0.26}$ for the Hubble+Pantheon dataset. In
this case, $\omega _{0}$ represents the current value of the EoS parameter
for DE, which exhibits negative behavior and is situated in the quintessence
epoch.  In addition, we explored numerous cosmological parameters to assess
the model's feasibility. The deceleration parameter demonstrates that the
Cosmos depicted by our model transitions smoothly from the early
decelerated phase to the present accelerated phase. Further, the present
value of the deceleration parameter corresponding to the model parameter
values constrained by the Hubble+Pantheon dataset is $ q_{0}=-0.46_{-0.07}^{+0.16}$. The transition redshift is  $z_{tr}=0.83_{-0.33}^{+0.54}$ for the Hubble+Pantheon dataset.
Finally, Fig. \ref{Om} shows that the slope of the $Om(z)$ diagnostic parameter $%
Om(z)$ is negative. Thus, the model presented here behaves like the
Cosmos's quintessence model. The $f(Q)$ cosmological model under
consideration can be viewed as a highly promising alternative to the $\Lambda
$CDM model, and more research into its possibilities would be both
intriguing and important. A similar examination will be carried out in a
future publication.\newline

\section*{Acknowledgments}
This work was supported and funded by the Deanship of Scientific 
Research at Imam Mohammad Ibn Saud Islamic University (IMSIU) 
(grant number IMSIU-RG23008)

\textbf{Data availability} All generated data are included in this manuscript.\newline

\end{document}